\documentclass[a4paper, 11pt]{article}
\pdfoutput=1
\usepackage{jheppubnew}
\usepackage{graphicx}
\usepackage{bm}
\usepackage{ifthen}
\usepackage{amsmath}
\usepackage{amsfonts}
\usepackage{amssymb}
\usepackage{amsthm}
\usepackage{mathrsfs}
\usepackage[all]{xy}
\usepackage{comment}
\usepackage[numbers,sort&compress]{natbib}
 \usepackage[utf8]{inputenc}

\newcommand{\bea}{\begin{eqnarray}\label}
\newcommand{\eea}{\end{eqnarray}}

\newcommand{\ang}[1]{\langle #1\rangle}

\newcommand{\be}{\begin{equation}}
\newcommand{\ee}{\end{equation}}
\renewcommand{\[}{\begin{equation}}
\renewcommand{\]}{\end{equation}}

\renewcommand{\O}{\mathcal{O}}

\newcommand{\ep}{\epsilon}

\renewcommand{\>}{\rangle}

\newcommand{\nn}{\nonumber}

\newcommand{\bs}[1]{\boldsymbol{#1}}

\newcommand{\ThreePt}{\empty}
\newcommand{\3}[1]{C_{
		\ifthenelse{\equal{\ThreePt}{\empty}}{#1}{
			\ifthenelse{\equal{#1}{\empty}}{\ThreePt}{\ThreePt,#1}}}}
\newcommand{\redef}[1]{{C'}_{
		\ifthenelse{\equal{\ThreePt}{\empty}}{#1}{
			\ifthenelse{\equal{#1}{\empty}}{\ThreePt}{\ThreePt,#1}}}}
\newcommand{\ren}[1]{C_{
		\ifthenelse{\equal{\ThreePt}{\empty}}{#1}{
			\ifthenelse{\equal{#1}{\empty}}{\ThreePt}{\ThreePt,#1}}}}
\newcommand{\sd}[1]{D_{
		\ifthenelse{\equal{\ThreePt}{\empty}}{#1}{
			\ifthenelse{\equal{#1}{\empty}}{\ThreePt}{\ThreePt,#1}}}}

\renewcommand{\vec}{\bs}

\title{Double Copy Structure of CFT Correlators}
\author[a]{Joseph A. Farrow,}
\author[a]{Arthur E. Lipstein}
\author[b]{and Paul McFadden.}
\affiliation[a]{Department of Mathematical Sciences, Durham University, Durham, DH1 3LE, U.K.}
\affiliation[b]{School of Mathematics, Statistics \& Physics, Newcastle University, Newcastle,  NE1 7RU, U.K.}
\emailAdd{joseph.a.farrow@durham.ac.uk}
\emailAdd{arthur.e.lipstein@durham.ac.uk}
\emailAdd{paul.l.mcfadden@newcastle.ac.uk}

\begin{document}

\abstract{

We consider the momentum-space 3-point correlators of currents, stress tensors and marginal scalar operators in general odd-dimensional conformal field theories.  We show that the flat space limit of these correlators is spanned by gauge and gravitational scattering amplitudes in one higher dimension which are related by a double copy. Moreover, we recast three-dimensional CFT correlators in terms of  tree-level Feynman diagrams without energy conservation, suggesting double copy structure beyond the flat space limit.

} 

\maketitle

\section{Introduction}

Scattering amplitudes and correlation functions are 
the basic ingredients from which many  important observables in quantum field theory are constructed.
From a theoretical standpoint, they are intimately related and exhibit remarkable mathematical properties. 
This connection is particularly transparent in a holographic setting, where CFT correlators can be expressed as Witten diagrams which reduce to scattering amplitudes in a suitable flat space limit  \cite{Penedones:2010ue}.  
Both scattering amplitudes and correlators can moreover be computed from lower-point objects.
For amplitudes, this takes the form of BCFW recursion \cite{Britto:2005fq} and unitarity methods \cite{Bern:1994zx,Bern:1994cg}, while 
for CFT correlators this  program
is known as the conformal bootstrap \cite{Ferrara:1973yt,Ferrara:1973vz,Polyakov:1974gs,Rattazzi:2008pe}.

In this paper, our goal is to make further contact between scattering amplitudes and correlation functions in order to reveal new mathematical structures and develop 
more efficient computational methods.  
One of the most remarkable properties of scattering amplitudes is a set of relations -- known collectively as the {\it double copy} -- enabling gravitational amplitudes to be expressed as a product of gauge theory amplitudes.
This was first observed in the context of tree-level string theory via the KLT relations between open and closed string amplitudes 
\cite{Kawai:1985xq}. In the field theory limit, these relations can then be extended to loop level using color-kinematics duality \cite{Bern:2008qj,Bern:2010ue}. Moreover, the double copy can be made manifest using worldsheet formulations \cite{Cachazo:2013hca,Mason:2013sva} which apply to a broad range of quantum field theories \cite{Cachazo:2014xea}.

In this paper, we will study CFT correlators in momentum space, 
where a direct connection to scattering amplitudes has been established \cite{Maldacena:2011nz, Raju:2012zr}.\footnote{The correspondence is also well understood  in Mellin space, see, {\it e.g.,} \cite{Mack:2009gy,Mack:2009mi,Penedones:2010ue, Fitzpatrick:2011ia}.}  
As we will see, CFT correlators contain specific poles whose residues correspond to scattering amplitudes in a bulk spacetime of one higher dimension.  If the CFT is Euclidean, this bulk spacetime is 
Lorentzian de Sitter enabling purely spatial boundary momenta to be lifted to null momenta in the bulk.  
This formulation is particularly relevant for the study of inflationary non-Gaussianities, where scattering amplitudes can be obtained from the flat space limit of cosmological in-in correlators, which are in turn  related holographically to boundary CFT correlators \cite{Maldacena:2002vr, McFadden:2010vh,McFadden:2011kk,Schalm:2012pi, Bzowski:2012ih,Mata:2012bx, Anninos:2014lwa, Isono:2016yyj}.  
Recent discussions 
at the level of the 4-point function include \cite{Ghosh:2014kba,Arkani-Hamed:2015bza,Arkani-Hamed:2018kmz,Arkani-Hamed:2017fdk,Arkani-Hamed:2018bjr,Benincasa:2018ssx}.  
Moreover,  KLT-like relations for the inflationary graviton 4-point function were identified in \cite{Li:2018wkt}. 
Alternatively, the connection with scattering amplitudes can be used to compute higher-point AdS Witten diagrams via an analogue of BCFW recursion \cite{Raju:2012zr, Raju:2011mp, Raju:2010by,Raju:2012zs,Albayrak:2018tam}. 

Here, our calculations will build on new results for CFT 
correlators of stress tensors, currents and scalars in general dimensions obtained by solving the conformal Ward identities in momentum space  \cite{Bzowski:2013sza,Bzowski:2015pba,Bzowski:2015yxv,Bzowski:2017poo,Bzowski:2018fql}. 
Our main result will be that 3-point correlators in general CFTs encode double copy relations familiar from scattering amplitudes. 
The origin of this connection can be understood by
 dressing the correlators with polarization vectors and interpreting them as cosmological correlators evaluated on the future boundary of de Sitter.  Scattering amplitudes then follow by taking the flat space limit, defined as the limit in which energy is conserved.  
Since energy is not conserved for cosmological correlators, to take this limit in practice requires analytic continuation of the momenta.  
In odd dimensions,  3-point CFT correlators  develop poles in this limit, and flat space scattering amplitudes can be read off from the most singular terms. In even dimensions, the flat space limit is more complicated to evaluate since the analytic structure of CFT correlators is more involved, so here we
will focus our attention on odd-dimensional cases. 
For a  parity-invariant but otherwise general 
CFT with $d>3$, we then find that the flat space limit of current correlators is spanned by scattering amplitudes in ordinary and higher-derivative Yang-Mills theory, while the flat space limit of stress tensor correlators is spanned by amplitudes in Einstein, $\phi R^2$ and Weyl-cubed gravity, where $\phi R^2$ is a curvature-squared theory coupled to scalars which reduces to a certain non-minimal conformal gravity in four dimensions \cite{Johansson:2018ues}. Remarkably, these theories are related via a double copy \cite{Broedel:2012rc,Johansson:2017srf}. 

In $d=3$, we show that the $\phi R^2$ contribution to the stress tensor 3-point function vanishes as a result of certain tensorial degeneracies.  
From the perspective of the double copy, this can be understood by writing the ordinary and higher-derivative Yang-Mills amplitudes in four dimensions in terms of spinor variables, whereupon the product corresponding to the 3-point graviton $\phi R^2$ amplitude vanishes. 
Instead, one can construct a non-vanishing product corresponding a graviton-graviton-scalar amplitude in the $\phi R^2$ theory, which arises as the flat space limit of a CFT correlator for two stress tensors and a marginal scalar. Furthermore, when this amplitude is written in terms of momenta and polarization vectors, it can be written as the square of a Yang-Mills-dilaton amplitude which we show arises in the flat space limit of correlators with two currents and a marginal scalar in general dimensions. Another special feature of $d=3$ CFTs is that the corresponding de Sitter correlators can be interpreted as {\it flat-space} tree-level Feynman diagrams, without energy conservation, 
dressed by conformal time integrals.  
As we will see, this form of the CFT correlators exhibits double copy structure beyond the flat space limit.

In $d=3$, CFT correlators  can conveniently be written in a spinor helicity formalism  \cite{Maldacena:2011nz}. This formalism has also recently been used to study 3-point correlators of higher spin currents in \cite{Nagaraj:2018nxq,Skvortsov:2018uru}. Along with this paper, we include the {\sc{Mathematica}} file {\tt{dSDoubleCopy.nb}} which has a comprehensive set of  multi-purpose functions for working with spinor helicity notation on dS$_4$. It provides specific algorithms for working with any 3-point expression analytically, as well as functions for the  numerical evaluation of $n$-point expressions.  Also included are multi-purpose functions for working with 3-point correlation functions in terms of polarization vectors and momenta.

The structure of this paper is as follows. In section \ref{amps}, we review some basic properties of scattering amplitudes. In sections \ref{dg3} and \ref{3d}, we analyze the flat space limit of correlators in $d>3$ and $d=3$ respectively. Finally, in section \ref{conclusion} we summarize our results and discuss open questions.  In Appendix \ref{wittendiagrams}, we compute de Sitter correlators corresponding to  $d=3$ CFTs and their reduction to flat space amplitudes.  In Appendix \ref{spinors} we describe the  spinor helicity formalism for 3d CFT correlators and present various useful formulae. 

\section{Amplitudes} \label{amps}

We begin with a review of some basic facts about scattering amplitudes that will be relevant later on.  In $(d+1)$-dimensional Minkowski space,\footnote{Our signature convention is mostly plus.} amplitudes can be written in terms of momenta $p^{\mu}_i$ and polarization vectors $\epsilon^{\mu}_i$ of the external particles labelled by the index $i$, where $\mu=0,1,...,d$. For massless external particles, the momenta and polarization vectors satisfy
\[
p_i\cdot p_i=0,\qquad \epsilon_i\cdot \epsilon_i=0,\qquad \epsilon_i\cdot p_i=0.
\]
If the particles have spin two, it is convenient to write their polarization tensors in terms of polarization vectors as $\epsilon_i^{\mu \nu}= \epsilon_i^{\mu} \epsilon_i^{\nu}$. 
Moreover, the momenta associated with an $n$-point amplitude are conserved,
\[
\sum_{i=1}^{n}p_{i}^{\mu}=0.
\]
In $(3+1)$-dimensions specifically, null momenta and polarizations can be written (suppressing the label $i$ for clarity) in terms of 2-component spinors as 
\begin{equation}
p^{\alpha\dot{\alpha}}=\lambda^{\alpha}\tilde{\lambda}^{\dot{\alpha}},\qquad \epsilon_{+}^{\alpha\dot{\alpha}}=\frac{\mu^{\alpha}\tilde{\lambda}^{\dot{\alpha}}}{\vphantom{\int_a^b}\langle \lambda\mu\rangle },\qquad \epsilon_{-}^{\alpha\dot{\alpha}}=\frac{\lambda^{\alpha}\tilde{\mu}^{\dot{\alpha}}}{\vphantom{\int_a^b}[\tilde{\lambda}\tilde{\mu}]},
\label{4dspinor}
\end{equation}
where the indices $\alpha,\dot{\alpha}$ label the fundamental representation of the Lorentz group which is locally equivalent to $SU(2) \times SU(2)$, $\left\{ \mu^{\alpha},\tilde{\mu}^{\dot{\alpha}}\right\} $ are arbitrary reference spinors encoding gauge invariance, and the brackets are defined by
\[
\left\langle ij\right\rangle =\lambda_{i}^{\alpha}\lambda_{j}^{\beta}\epsilon_{\alpha\beta},\qquad \left[ij\right]=\tilde{\lambda}_{i}^{\dot{\alpha}}\tilde{\lambda}_{j}^{\dot{\beta}}\epsilon_{\dot{\alpha}\dot{\beta}},
\]  
where $\epsilon$ is the antisymmetric Levi-Civita symbol.

In this paper, we will be primarily interested in two gauge theories, ordinary Yang-Mills (YM) and a higher-derivative Yang-Mills theory with an $F^3$ interaction (where $F$ is the Yang-Mills field strength) which was recently constructed in \cite{Johansson:2017srf}. Using the double copy, the amplitudes of these two gauge theories can be combined to obtain amplitudes in Einstein (EG), $\phi R^2$, and Weyl-cubed ($W^3$) gravity, where $\phi R^2$ refers to a curvature squared theory coupled to scalars \cite{Broedel:2012rc}. Note that the $F^3$ theory constructed in \cite{Johansson:2017srf} contains both gluons and scalars. Moreover, after performing the double copy, the resulting theories will contain fields other than gravitons. For example, the double copy of Yang-Mills corresponds to Einstein gravity coupled to a dilaton and 2-form gauge field (which can be dualized to an axion in four dimensions). 

The double copy is most easily seen at the level of 3-point graviton and gluon amplitudes in these theories, which are given by
\begin{equation}
\mathcal{A}_{EG}=(\mathcal{A}_{YM})^{2},\qquad \mathcal{A}_{\phi R^{2}}^{222}=\mathcal{A}_{F^{3}}\mathcal{A}_{YM},\qquad \mathcal{A}_{W^{3}}=(\mathcal{A}_{F^{3}})^{2},
\label{gravityamp}
\end{equation}
where
\begin{equation}
\mathcal{A}_{YM}=\epsilon_{1}\cdot\epsilon_{2}\,\,\epsilon_{3}\cdot p_{1}+\mathrm{cyclic},\qquad \mathcal{A}_{F^{3}}=\epsilon_{1}\cdot p_{2}\,\,\epsilon_{2}\cdot p_{3}\,\,\epsilon_{3}\cdot p_{1}.
\label{gaugeamp}
\end{equation}
We use the superscript $222$ on the $\mathcal{A}_{\phi R^{2}}$ to emphasize that it is a graviton amplitude. We will also consider graviton-graviton-scalar amplitudes in this theory, which will have the superscript $220$. These formulae can be naturally extended to higher points using worldsheet formulations. In particular, worldsheet formulae for Yang-Mills and Einstein gravity were initially proposed in \cite{Cachazo:2013hca}, and were later extended to $F^3$, $\phi R^2$ and $W^3$ in \cite{Mason:2013sva,He:2016iqi,Azevedo:2017lkz}. Although we will only need 3-point amplitudes for the purposes of this paper, we expect that worldsheet formulae will be useful for extending our results to higher points.

The double copy structure described above can be also anticipated using bosonic string theory. In particular, open string theory predicts the 3-point gluon amplitude
\[
\mathcal{A}_{YM}+ \alpha' \mathcal{A}_{F^3}
\]
and closed string theory predicts the 3-point graviton amplitude
\[
\left(\mathcal{A}_{YM}+\ensuremath{\alpha'}\mathcal{A}_{F^{3}}\right)^{2}=\mathcal{A}_{EG}+2\ensuremath{\alpha'}\mathcal{A}_{\phi R^{2}}^{222}+\alpha'^{2}\mathcal{A}_{W^{3}},
\]
where $\alpha'$ is related to the square of the string length. This result was recently generalized to any number of points in \cite{Azevedo:2018dgo}, where it was shown that tree-level open bosonic string amplitudes can be decomposed into $YM+F^3$ amplitudes times a basis of worldsheet integrals encoding the $\alpha '$ dependence, from which closed bosonic string amplitudes can be obtained via a double copy.

Before closing this section, let us mention that in four dimensions, the $\phi R^2$ theory reduces to a certain non-minimal conformal gravity whose Lagrangian was recently constructed in \cite{Johansson:2018ues}. Moroever, the 3-point graviton amplitudes in this theory vanish. This is most easily seen from the double copy using spinor helicity notation. In particular, we have 
\[
\mathcal{A}_{F^{3}}(1^{-},2^{-},3^{-})=\left\langle 12\right\rangle \left\langle 23\right\rangle \left\langle 31\right\rangle ,\qquad \mathcal{A}_{YM}(1^{-},2^{-},3^{+})=\frac{\left\langle 12\right\rangle ^{3}}{\left\langle 23\right\rangle \left\langle 31\right\rangle },
\]
where the superscripts denote the helicity of the external particles. Moreover, we have
\[
\mathcal{A}_{F^{3}}(1^{-},2^{-},3^{+})=\mathcal{A}_{YM}(1^{-},2^{-},3^{-})=0,
\]
with opposite helicity amplitudes obtained by complex conjugation. The 3-point graviton amplitudes are then manifestly zero, {\it e.g.,} \[
\mathcal{A}_{\phi R^{2}}^{222}(1^{--},2^{--},3^{--})=\mathcal{A}_{F^3}(1^-,2^-,3^-)\mathcal{A}_{YM}(1^-,2^-,3^-)=0. 
\]
If instead we take the product of the two {\it nonzero} amplitudes, we obtain an amplitude for two gravitons and a scalar which arises in non-minimal conformal gravity:
\begin{equation}
\mathcal{A}_{\phi R^{2}}^{220}(1^{--},2^{--},3^{-+})=\mathcal{A}_{F^{3}}(1^{-},2^{-},3^{-})\mathcal{A}_{YM}(1^{-},2^{-},3^{+})=\left\langle 12\right\rangle ^{4}.
\label{phiw2}
\end{equation}
In terms of polarization vectors, this corresponds to
\begin{equation}
\mathcal{A}_{\phi R^{2}}^{220}=\left(\epsilon_{1}\cdot p_{2}\,\epsilon_{2}\cdot p_{1}\right)^{2},
\label{phir2vec}
\end{equation}
which can be obtained from $\mathcal{A}_{\phi R^{2}}^{222}$ in $d>3$ by making the replacement $\epsilon_{3}^{\mu}\epsilon_{3}^{\nu}\rightarrow\eta^{\mu\nu}$, where $\eta_{\mu\nu}$ is the Minkowski metric. Note that the graviton-scalar-scalar amplitude obtained from the product $\mathcal{A}_{F^{3}}(1^{-},2^{-},3^{-})\mathcal{A}_{YM}(1^{+},2^{+},3^{-})$ vanishes by momentum conservation. Hence, $\mathcal{A}_{\phi R^{2}}^{220}$ is the only 3-point amplitude which can be obtained from combining Yang-Mills and the $F^3$ gluon amplitudes in four dimensions. Generalizations to higher points in the form of worldsheet formulas can be found in \cite{Berkovits:2004jj,Dolan:2008gc,Adamo:2012xe,Farrow:2018yqf}.

In general dimensions, the graviton-graviton-scalar amplitude in \eqref{phir2vec} can be written as the square of a gluon-gluon-scalar amplitude arising from a $\phi F^2$ interaction,
\[
\mathcal{A}_{\phi R^{2}}^{220}=\left(\mathcal{A}_{\phi F^{2}}\right)^{2},
\label{ymdc}
\]
where
\begin{equation}
\mathcal{A}_{\phi F^{2}}=\epsilon_{1}\cdot p_{2}\,\epsilon_{2}\cdot p_{1}.
\label{dilaton}
\end{equation}
To our knowledge, the double copy of Yang-Mills-dilaton amplitudes has not been previously considered, so it would be interesting to see if the relation in \eqref{ymdc} extends to higher points.

\section{\texorpdfstring{CFT correlators for  $\bs{d > 3}$ }{CFT correlators for  d>3}}  \label{dg3}

Let us now shift our attention to $d$-dimensional Euclidean CFT correlators.   In this section we consider all odd $d>3$, and in the following section we will return to the case $d=3$.  

Since we are interested in connecting to scattering amplitudes, we will focus on the transverse-traceless parts of CFT correlators.\footnote{The remaining non-transverse traceless pieces are fixed in terms of lower-point functions by the trace and transverse Ward identities, and can be completely reconstructed from these identities,
see \cite{Bzowski:2013sza, Bzowski:2017poo, Bzowski:2018fql}.}
As discussed in \cite{Bzowski:2013sza}, 3-point functions can be decomposed in a minimal basis of transverse-traceless tensors constructed from the metric and the momenta $\vec{p}_i$, where $i=1,2, 3$ labels the insertion.   Each basis tensor in this decomposition appears with a corresponding form factor which is a scalar function of the momentum magnitudes 
\[
p_i=+\sqrt{\vec{p}_i^{2}}.
\]   
For 3-point functions, any contraction of momenta can be expressed in terms of the momentum magnitudes via momentum conservation,
\[\label{CFTmomcons}
\vec{p}_1+\vec{p}_2+\vec{p}_3 = 0,
\]
for example $\vec{p}_1\cdot\vec{p}_2=(p_3^2-p_1^2-p_2^2)/2$.
We emphasize that these momenta are those of a $d$-dimensional Euclidean CFT and hence are not in general null.  For physical kinematics, all momentum magnitudes are instead such that $p_i\ge 0$ and the triangle inequalities are satisfied ({\it i.e.,} $p_1+p_2\ge p_3$).
The null momenta relevant for $(d+1)$-dimensional scattering amplitudes, as discussed in the previous section, are related to those of the $d$-dimensional Euclidean CFT by 
\[\label{bulkpdef}
p_i^{\mu}=\left(p_i,\, \vec{p}_i\right),
\]
where the first component corresponds to the time direction.
While these $p_i^\mu$ are null, the total `energy' defined by
\[\label{Edef}
E=p_1+p_2+p_3
\]
is non-vanishing for $p_i$ derived from Euclidean 3-point functions with physical kinematics.
Nevertheless, we can reach configurations with $E=0$, for which energy is conserved from the perspective of $(d+1)$-dimensional flat space scattering amplitudes, by a suitable analytic continuation.  As we will see,  CFT correlators develop poles in the flat space limit $E\rightarrow 0$, whose leading coefficients correspond to these scattering amplitudes.

To make the relation to scattering amplitudes more transparent, we will contract all Lorentz indices on CFT correlators with polarization vectors satisfying
\[
\vec{\epsilon}_i\cdot\vec{\epsilon}_i=0,\qquad \vec{\epsilon}_i\cdot\vec{p}_i=0.
\]
These $d$-dimensional polarization vectors are then related to those of the $(d+1)$-dimensional scattering amplitudes by
\[\label{epdef}
\epsilon^{\mu}_i=(0,\vec{\epsilon}_i).
\]
With the tensorial structure of correlators thus dealt with, the remaining scalar form factors can all be expressed as linear combinations of `triple-$K$' integrals
\begin{align}
I_{\alpha \{ \beta_1, \beta_2, \beta_3 \}}(p_1, p_2, p_3) & = \int_0^\infty \mathrm{d} x \: x^\alpha \prod_{i=1}^3 p_i^{\beta_i} K_{\beta_i}(p_i x), \label{tripleK} 
\end{align}
where $K_{\beta_i}$ is a modified Bessel function of the second kind.
The exact linear combinations and arguments $\alpha$, $\beta_i$ appearing can be found by solving the relevant conformal Ward identities, as detailed in \cite{Bzowski:2013sza}.   In order to write compact expressions valid for general $d$, it is useful to further define the `reduced' triple-$K$ integral 
\begin{align}
J_{N \{ k_1,k_2,k_3 \}} & = I_{\frac{d}{2} - 1 + N \{ \Delta_1 - \frac{d}{2} + k_1, \Delta_2 - \frac{d}{2} + k_2 , \Delta_3 - \frac{d}{2} + k_3  \}}, \label{a:J}
\end{align}
where the $N$ and $k_i$ are integer arguments and $\Delta_i$ is the conformal dimension of the $i$-th operator.  For conserved currents $\Delta_i=d-1$, while for stress tensors and marginal scalars $\Delta_i=d$. 
In addition, we will use the symmetric polynomials 
\[\label{c123def}
a_{ij}=p_{i}+p_{j},\quad b_{ij}=p_{i}p_{j},\quad b_{123}=p_{1}p_{2}+p_{2}p_{3}+p_{3}p_{1},\quad c_{123}=p_{1}p_{2}p_{3},
\]
and the quantity
\[
J^{2}=E\,\left(p_{1}+p_{2}-p_{3}\right)\left(p_{1}-p_{2}+p_{3}\right)\left(-p_{1}+p_{2}+p_{3}\right),
\]
with $E$ given by \eqref{Edef}.
By Heron's formula, $\sqrt{J^2}/4$ is the area of the triangle with side lengths given by the $p_i$.   For physical kinematics, the triangle inequality ensures that $J^2\ge 0$, vanishing only for collinear momenta, though this no longer the case after analytic continuation of the $p_i$.

To extract scattering amplitudes from the CFT correlators, we need to evaluate the leading behaviour of the triple-$K$ integral \eqref{tripleK} in the flat space limit $E\rightarrow 0$.
For $\alpha > 1/2$, as will always be the case, this leading behaviour is 
\begin{align}
\lim_{E \rightarrow 0} I_{\alpha\{\beta_1, \beta_2, \beta_3\}} \rightarrow (\pi/2)^{3/2} \,\Gamma (\alpha - 1/2)\, p_1^{\beta_1 - 1/2} p_2^{\beta_2 - 1/2} p_3^{\beta_3 - 1/2} \,E^{1/2 - \alpha}.
\label{flatspacetriplek}
\end{align}  
For half-integer $\beta_i$, such as arise for the odd-dimensional correlators of interest here, this formula can be proved by noting that 
\[
p_i^{\beta_i}K_{\beta_i}(p_i x)=(\pi/2)^{1/2} x^{-\beta_i} f(p_i x)e^{-p_ix},
\]
where $f(p_i x)$ is a polynomial whose highest term is $(p_i x)^{\beta_i-1/2}$.  The triple-$K$ integral \eqref{tripleK} can now be evaluated as a sum of Euler gamma functions, and the leading term as $E\rightarrow 0$ is that associated with the highest power of $x$.

In fact, we believe  \eqref{flatspacetriplek} holds for fully general $\beta_i$ since the leading behavior as $E\rightarrow 0$ derives from the asymptotic behavior  $\ensuremath{K_{\beta_i}(p_ix)\rightarrow\sqrt{\pi/(2p_ix)}e^{-p_ix}}$ of the Bessel function as $x\rightarrow\infty$, which yields precisely  \eqref{flatspacetriplek}.   
However, while odd-dimensional correlators are rational functions, in even dimensions correlators have a more complicated analytic structure.  A careful specification of the analytic continuations needed to reach $E\rightarrow 0$ is then required.\footnote{In even dimensions one also encounters divergences  necessitating regularization and renormalization \cite{Bzowski:2017poo,Bzowski:2018fql}.  
As these divergences arise from the lower limit $x\rightarrow 0$ of the triple-$K$ integral, while the behavior as $E\rightarrow 0$ derives from the upper limit $x\rightarrow \infty$, we expect these effects can be cleanly disentangled.}
In $d=4$, for example, the relevant triple-$K$ integrals can all be derived from a single master integral $I_{1\{000\}}$ which can be evaluated in terms of dilogarithms \cite{Bzowski:2015pba, Bzowski:2017poo}.    This integral is equivalent to a 1-loop triangle diagram in flat space whose analytic structure has been studied in  \cite{tHooft:1978xw,Davydychev:1995mq, Chavez:2012kn, Maldacena:2015iua}.  
We hope to analyze this further in future work.

In the following, we now proceed to evaluate the flat space limit $E\rightarrow 0$ for odd-dimensional 3-point correlators of currents, stress tensors and marginal scalars. 
We find the resulting scattering amplitudes are spanned by the gauge and gravitational theories described in the previous section, implying that the correlators encode a double copy structure. 

\subsection{\texorpdfstring{${\<JJJ\>}$}{<JJJ>} }
First, we consider the 3-point function of conserved currents for general odd dimensions $d>3$.  Taking the general solution of the conformal Ward identities given in \cite{Bzowski:2013sza, Bzowski:2017poo} and contracting with polarization vectors, we find
\begin{equation}
\left\langle J J J \right\rangle =A_{1}(p_{1},p_{2},p_{3})\,{\epsilon}_{1}\cdot {p}_{2}\,{\epsilon}_{2}\cdot {p}_{3}\, \epsilon_{3}\cdot p_{1}+\big[A_{2}(p_{1},p_{2},p_{3})\,\epsilon_{1}\cdot\epsilon_{2}\,\epsilon_{3}\cdot p_{1}+\mathrm{cyclic}\big]
\label{jjj}
\end{equation}
where the form factors are given by
\begin{align}
A_1 = C_1 J_{3 \{000\}}, \qquad 
A_2  = C_1 J_{2 \{001\}} + C_2 J_{1 \{000\}},
\end{align}
where $C_1$ and $C_2$ are constants
and the reduced triple-$K$ integrals are defined in \eqref{a:J}. 
For clarity, we suppress color factors and the overall delta function enforcing momentum conservation.
To connect with bulk scattering amplitudes, we have used \eqref{bulkpdef} and \eqref{epdef} to replace $\bs{\epsilon}_i\cdot \bs{p}_j = \epsilon_i\cdot p_j$.

The conformal Ward identities further relate $C_2$ to the normalization $C_{JJ}$ of the current 2-point function via
\[\label{C2JJJ}
C_2 = \# C_1 + \# C_{JJ},
\]
where the $\#$ are dimension-dependent coefficients whose form will not be important here.\footnote{See (3.33) of  \cite{Bzowski:2017poo} for details.} 
In the flat space limit, using \eqref{a:J} and \eqref{flatspacetriplek}, we find \eqref{jjj} reduces to
\begin{equation}
\ensuremath{\lim_{E\rightarrow0}\left\langle JJJ\right\rangle }\propto c_{123}^{(d-3)/2}\left[\frac{C_{1}}{E^{(d+3)/2}}\left(\mathcal{A}_{F^{3}}+\mathcal{O}(E)\right)+\frac{C_{JJ}}{E^{(d-1)/2}}\left(\mathcal{A}_{YM}+\mathcal{O}(E)\right)\right],
\label{flatjjj}
\end{equation}
where the  amplitudes $\mathcal{A}_{YM}$ and $\mathcal{A}_{F^{3}}$ are defined in \eqref{gaugeamp}, and the symmetric polynomial $c_{123}$ is given in \eqref{c123def}.  We used \eqref{C2JJJ} to replace $C_2$ with $C_{JJ}$, noting that the term proportional to $C_1$ yields a contribution that is subleading in the flat space limit.  We have also allowed a rescaling of the arbitrary constant $C_1$. 

We thus find that the flat space limit of the current correlator in general CFTs is spanned by ordinary and higher-derivative Yang-Mills amplitudes:
\begin{align}
\lim_{E\rightarrow0}\frac{E^{(d+3)/2}}{c_{123}^{(d-3)/2}}\left.\left\langle JJJ\right\rangle\right|_{C_{JJ}=0}\propto \mathcal{A}_{F^{3}}, \\[2ex]
\lim_{E\rightarrow0}\frac{E^{(d-1)/2}}{c_{123}^{(d-3)/2}}\left.\left\langle JJJ\right\rangle \right|_{C_{1}=0}\propto \mathcal{A}_{YM}. 
\end{align}
\vspace{1ex}

\subsection{\texorpdfstring{$\left\langle TTT\right\rangle$}{<TTT>} }

Next we consider the stress tensor 3-point function for general odd dimensions $d>3$.  After contracting with polarization vectors,  the general solution to the conformal Ward identities takes the form  \cite{Bzowski:2013sza, Bzowski:2017poo} 
\begin{align}
\left\langle T T T \right\rangle  
&\nonumber =A_{1}(p_1,p_2,p_3)\left(\epsilon_{1}\cdot p_{2}\,\epsilon_{2}\cdot p_{3}\,\epsilon_{3}\cdot p_{1}\right)^{2}  \\ 
\nonumber &\quad +\big(A_{2}(p_1,p_2,p_3)\,\epsilon_{1}\cdot\epsilon_{2}\,\epsilon_{1}\cdot p_{2}\,\epsilon_{2}\cdot p_{3}\left(\epsilon_{3}\cdot p_{1}\right)^{2} + \mathrm{cyclic}\big)\\
\nonumber &\quad +\big(A_{3}(p_1,p_2,p_3)\left(\epsilon_{1}\cdot\epsilon_{2}\right)^{2}\left(p_{1}\cdot\epsilon_{3}\right)^{2}+\mathrm{cyclic}\big)\\
\nonumber &\quad +\big(A_{4}(p_1,p_2,p_3)\,\epsilon_{1}\cdot\epsilon_{3}\,\epsilon_{2}\cdot\epsilon_{3}\,\epsilon_{1}\cdot p_{2}\,\epsilon_{2}\cdot p_{3}+\mathrm{cyclic}\big)\\
&\quad +A_{5}(p_1,p_2,p_3)\,\epsilon_{1}\cdot\epsilon_{2}\,\epsilon_{2}\cdot\epsilon_{3}\,\epsilon_{3}\cdot\epsilon_{1},
\label{tttgeneral} 
\end{align}
where the form factors are  
\begin{align}\label{TTTformfactorsgend}
A_1 & = C_1 J_{6 \{000\}}, \\
A_2 & = 4 C_1 J_{5 \{001\}} + C_2 J_{4 \{000\}}, \\
A_3 & = 2 C_1 J_{4 \{002\}} + C_2 J_{3 \{001\}} + C_3 J_{2 \{000\}},\\
A_4 & = 8 C_1 J_{4 \{110\}} - 2 C_2 J_{3 \{001\}} + C_4 J_{2 \{000\}}, \\
A_5 & = 8 C_1 J_{3 \{111\}} + 2 C_2 \left( J_{2 \{110\}} + J_{2 \{101\}} + J_{2 \{011\}} \right) + C_5 J_{0 \{000\}}.
\end{align}
The reduced triple-$K$ integrals are given in \eqref{a:J}, and  
 the cyclic permutations in \eqref{tttgeneral} act on all polarization vectors and momenta, as well as the arguments of each form factor.
To avoid confusion, we emphasize that the constants $C_n$ associated with the different correlators are unrelated: our $C_n$ here is simply a shorthand for $C_n^{(TTT)}$, and so forth.


The conformal Ward identities further impose that only three of the five constants in \eqref{TTTformfactorsgend} are linearly independent.\footnote{See section 3.4.2 of \cite{Bzowski:2017poo} for details.} 
Schematically,
\[ 
C_{4}=2C_{3}+\#C_{2},\quad C_{5}=\#C_{1}+\#C_{2}+\#C_{3},
\]
where the coefficients denoted by $\#$ are dimension-dependent and will not be needed for our analysis as the corresponding terms are subleading in the flat space limit.  
One can similarly replace $C_3$ with the 2-point normalization $C_{TT}$ since
\[
C_3 = \# C_1 + \# C_2 + \# C_{TT},
\]
and the terms proportional to $C_1$ and $C_2$ yield subleading contributions.

Applying \eqref{flatspacetriplek}, we then find that
\begin{align}
\lim_{E\rightarrow0}\ensuremath{\left\langle TTT\right\rangle } &\propto c_{123}^{(d-1)/2}\left[\frac{C_{1}}{E^{(d+9)/2}}\left(\mathcal{A}_{W^{3}}+\mathcal{O}(E)\right)\right.\nn\\&\qquad\qquad\quad\quad
\left.+\frac{C_{2}}{E^{(d+5)/2}}\left(\mathcal{A}_{\phi R^{2}}^{222}+\mathcal{O}(E)\right)+\frac{C_{TT}}{E^{(d+1)/2}}\left(\mathcal{A}_{EG}+\mathcal{O}(E)\right)\right],
\label{flatttt}
\end{align}
where the gravitational amplitudes $\mathcal{A}_{EG}$, $\mathcal{A}_{\phi R^{2}}^{222}$, $\mathcal{A}_{W^3}$ are defined in \eqref{gravityamp} and we have permitted independent rescalings of the constants $C_1$ and $C_2$.  The flat space limit of the stress tensor 3-point function in a general CFT is thus spanned by the Weyl-cubed, $\phi R^2$ and  Einstein gravity amplitudes:
\begin{align}\label{TTTflat1}
\lim_{E\rightarrow0}\frac{E^{(d+9)/2}}{c_{123}^{(d-1)/2}}\left.\left\langle TTT\right\rangle \right|_{C_{2}=C_{TT}=0}&\propto \mathcal{A}_{W^{3}},\\[2ex]\label{TTTflat2}
\lim_{E\rightarrow0}\frac{E^{(d+5)/2}}{c_{123}^{(d-1)/2}}\left.\left\langle TTT\right\rangle \right|_{C_{1}=C_{TT}=0}&\propto \mathcal{A}_{\phi R^{2}}^{222},\\[2ex]\label{TTTflat3}
\lim_{E\rightarrow0}\frac{E^{(d+1)/2}}{c_{123}^{(d-1)/2}}\left.\left\langle TTT\right\rangle \right|_{C_{1}=C_{2}=0}&\propto \mathcal{A}_{EG}.
\end{align}
Remarkably, these gravity amplitudes are related to the gauge theory amplitudes arising in the flat space limit of ${\left\langle JJJ\right\rangle }$ via the double copy relations \eqref{gravityamp}.

\subsection{\texorpdfstring{$\langle JJ\O\rangle$}{<JJO>}}
Next, let us look at the correlator of two currents and a marginal scalar operator. In this case, the general solution to the conformal Ward identities is \cite{Bzowski:2013sza, Bzowski:2018fql}
\begin{equation}
\left\langle JJ\O\right\rangle =-A_{1}(p_{1},p_{2},p_{3})\,{\epsilon}_{1}\cdot{p}_{2}\,{\epsilon}_{2}\cdot{p}_{1}+A_{2}(p_{1},p_{2},p_{3})\,\epsilon_{1}\cdot\epsilon_{2},
\label{jjogdcor}
\end{equation}
where the form factors are given by
\begin{align}
A_{1}=C_{1}J_{2\{000\}}, \qquad 
A_{2}=C_{1}J_{1\{001\}}+C_{2}J_{0\{000\}}.
\end{align}
The Ward identities further impose the relation 
\[
C_{2}=-\frac{1}{2}\Delta_{3}\left(\Delta_{3}-d+2\right)C_{1}=-C_{1}d
\]
where for a marginal scalar $\Delta_3=d$. Using \eqref{flatspacetriplek}, we obtain
\[\label{JJOflat1}
\lim_{E\rightarrow0}
\left\langle JJ\O\right\rangle \propto c_{123}^{(d-3)/2}p_3 \frac{C_1 }{E^{(d+1)/2}} \Big(\mathcal{A}_{\phi F^2}+\mathcal{O}(E)\Big).
\]
The flat space limit of $\langle JJ\O\rangle$ thus encodes the Yang-Mills-dilaton amplitude in \eqref{dilaton},  
\[\label{JJOflat2}
\lim_{E\rightarrow0} \frac{E^{(d+1)/2}}{c_{123}^{(d-3)/2}p_3}\left\langle JJ\O\right\rangle \propto \mathcal{A}_{\phi F^{2}}.
\]

\subsection{\texorpdfstring{$\langle TT\O\rangle$}{<TTO>}}
Finally, we consider the correlator of two stress tensors and a marginal scalar, given by \cite{Bzowski:2013sza, Bzowski:2018fql}
\begin{align}
\left\langle TT\O\right\rangle &=A_{1}(p_1,p_2,p_3)\,\left(\epsilon_{1}\cdot p_{2}\,\epsilon_{2}\cdot p_{1}\right)^{2}\nn\\&\quad  -A_{2}(p_1,p_2,p_3)\,\epsilon_{1}\cdot\epsilon_{2}\,\epsilon_{1}\cdot p_{2}\,\epsilon_{2}\cdot p_{1}+A_{3}(p_1,p_2,p_3)\left(\epsilon_{1}\cdot\epsilon_{2}\right)^{2},
\label{ttogdcor}
\end{align}
where 
\begin{align}
A_1 & = C_1 J_{4 \{000\}}, \\
A_2 & = 4 C_1 J_{3 \{001\}} + C_3 J_{2 \{000\}}, \\
A_3 & = 2 C_1 J_{2 \{002\}} + C_2 J_{1 \{001\}} + C_3 J_{0 \{000\}},
\end{align}
and the constants are related by
\begin{align}
C_{2}&=\left(\Delta_{3}+2\right)\left(d-\Delta_{3}-2\right)C_{1}=-2(d+2)C_{1},\\
C_{3}&=\frac{1}{4}\Delta_{3}\left(\Delta_{3}+2\right)\left(d-\Delta_{3}\right)C_{1}=0.
\end{align}
Using \eqref{flatspacetriplek}, we then obtain
\[
\lim_{E\rightarrow0}\left\langle TT\O\right\rangle \propto c_{123}^{(d-1)/2} \frac{C_{1}}{E^{(d+5)/2}}\left(\mathcal{A}_{\phi R^{2}}^{220}+\mathcal{O}\left(E\right)\right).
\]
The flat space limit thus encodes the amplitude in \eqref{phir2vec},   
\[
\lim_{E\rightarrow0}\frac{E^{(d+5)/2}}{c_{123}^{(d-1)/2}}\left\langle TT\O\right\rangle \propto\mathcal{A}_{\phi R^{2}}^{220}
\]
Furthermore, this is the square of the gauge theory amplitude arising in the flat space limit of $\langle JJ\O\rangle$ via \eqref{ymdc}.

To summarize, in general odd dimensions $d>3$ we have seen that CFT 
correlators of currents, stress tensors, and marginal scalars  exhibit a double copy structure via their flat space limit.   In the next section, we will extend this discussion to the case $d=3$.

\section{\texorpdfstring{CFT correlators for $\bs{d=3}$}{CFT correlators for d=3} } \label{3d}

In three dimensions, the form factors for $\<TTT\>$ exhibit degeneracies which leave the correlator invariant.  
These degeneracies arise since specific tensor structures in the form factor basis are equivalent up to terms derived from a 4-form.  In three dimensions, this 4-form vanishes as an index is necessarily repeated.  As a result, one obtains a set of equivalence relations between the form factors parametrized by three arbitrary functions of the momentum magnitudes.\footnote{See appendix A.5 of \cite{Bzowski:2017poo}.  Alternatively, one can view these degeneracies as arising from the existence of the cross-product in three dimensions, see appendix A.2 of \cite{Bzowski:2013sza}.  For a position space analysis, see \cite{Osborn:1993cr}.}   These relations eliminate a tensor structure such that the 3-point function depends on only two, rather than three, arbitrary constants.

Here, we show how to fix these degeneracies  so as to make the connection to scattering amplitudes manifest.  
The vanishing tensor structure in $\<TTT\>$ then corresponds to the vanishing 3-point graviton scattering amplitude in the $\phi R^2$ theory, which reduces to conformal gravity in four dimensions.
 The double copy moreover predicts a nonzero graviton-graviton-scalar amplitude in the $\phi R^2$ theory, which we show arises from the flat space limit of $\ang{TT\O}$ where $\O$ is a marginal scalar operator.  For this calculation, we also need to take into account a degeneracy of the form factor basis.

Another special feature of $d=3$ is that CFT correlators can be derived from 3-point functions in dS$_4$ which are equivalent to flat space Feynman diagrams without energy conservation, dressed by conformal time integrals.  
This connection allows us to re-write correlators in a manner suggesting the persistence of a  double copy structure beyond the flat space limit.
In the following we will point out these results where relevant, deferring all calculations to appendix  \ref{wittendiagrams}.

\subsection{\texorpdfstring{$\langle JJJ\rangle$}{<JJJ>}}
In $d=3$, the form factors for the current 3-point function are \cite{Bzowski:2013sza,Bzowski:2017poo}
\[
A_1 =\frac{2 C_1}{E^3}, \qquad 
A_2=\frac{C_1 p_3}{E^2} - \frac{2C_{JJ}} {E},
\]
where $C_{JJ}$ is the normalization of the 2-point function.\footnote{We suppress a factor relating to the  color and charge.}

Plugging this into \eqref{jjj} then gives
\[\label{F3full}
\left\langle JJJ\right\rangle =\frac{C_{1}}{E^{3}}\left(2\mathcal{A}_{F^{3}}+E\tilde{\mathcal{A}}_{YM}\right)-\frac{2C_{JJ}}{E}\mathcal{A}_{YM}
\]
where the new structure
\[\label{tildeAYMdef}
\tilde{\mathcal{A}}_{YM}=\left(\epsilon_{1}\cdot\epsilon_{2}\,\epsilon_{3}\cdot p_{1}\right)p_{3}+\mathrm{cyclic}
\]
resembles the YM amplitude but is not an actual scattering amplitude.

The term proportional to $C_1$ can be derived from a 3-point function for an $F^3$ interaction in dS$_4$.  As we show in appendix \ref{wittendiagrams}, this can be related to a flat space Feynman diagram without energy conservation, from which we find 
\begin{equation}
 2\mathcal{A}_{F^{3}}+E\tilde{\mathcal{A}}_{YM} \propto \mathcal{M}_{F^{3}}
\label{F3}
\end{equation}
where  
\[\label{MF3res}
\mathcal{M}_{F^{3}} \propto  
F_{\mu\nu}^{(1)}F_{\nu\rho}^{(2)}F_{\rho\mu}^{(3)},\qquad F_{\mu\nu}^{(\alpha)}=p_{[\mu}^{(\alpha)}\epsilon_{\nu]}^{(\alpha)}.
\]
Here, indices should be contracted using the $(3+1)$-dimensional Minkowski metric, with momenta and polarization vectors as given in \eqref{bulkpdef} and \eqref{epdef} with $(\alpha)$ labelling the insertion.

\subsection{\texorpdfstring{$\left\langle TTT\right\rangle$ }{<TTT>}}

Let us now turn to the stress tensor 3-point function.  In $d=3$, the  form factors are \cite{Bzowski:2013sza, Bzowski:2017poo}
\begin{align}
 A_1 &= \frac{8 C_1}{E^6} \left[ E^3 + 3 E b_{123} + 15 c_{123} \right], \\
\nn A_2 &= \frac{8 C_1}{E^5} \left[ 4 p_3^4 + 20 p_3^3 a_{12} + 4 p_3^2 (7 a_{12}^2 + 6 b_{12}) + 15 p_3 a_{12} (a_{12}^2 + b_{12}) + 3 a_{12}^2 (a_{12}^2 + b_{12}) \right]\\
& \qquad +  \frac{2 C_2}{E^4} \left[ E^3 + E b_{123} + 3 c_{123} \right], \\
\nn A_3 & = \frac{2 C_1 p_3^2}{E^4} \left[ 7 p_3^3 + 28 p_3^2 a_{12} + 3 p_3 (11 a_{12}^2 + 6 b_{12}) + 12 a_{12} ( a_{12}^2 + b_{12} ) \right]\\
 & \qquad + \: \frac{C_2 p_3^2}{E^3} \left[ p_3^2 + 3 p_3 a_{12} + 2 (a_{12}^2 + b_{12}) \right] - \frac{2 C_{TT}}{E^2} \left[ E^3 - E b_{123} - c_{123} \right], 
\end{align}
\begin{align}
 \nn A_4 & = \frac{4 C_1}{E^4} \left[ -3 p_3^5 - 12 p_3^4 a_{12} - 9 p_3^3 (a_{12}^2 + 2 b_{12}) + 9 p_3^2 a_{12} (a_{12}^2 - 3 b_{12}) \right. \\
\nonumber &\qquad\qquad \left. +  (4 p_3 + a_{12}) (3 a_{12}^4 - 3 a_{12}^2 b_{12} + 4 b_{12}^2) \right] \\
\nonumber & \qquad + \frac{C_2}{E^3} \left[ -p_3^4 - 3 p_3^3 a_{12} - 6 p_3^2 b_{12} + a_{12} (a_{12}^2 - b_{12}) (3 p_3 + a_{12}) \right] 
\\& \qquad
 - \frac{4 C_{TT}}{E^2} \left[ E^3 - E b_{123} - c_{123} \right], \\
\nn A_5 & = \frac{2 C_1}{E^3} \left[ -3 E^6 + 9 E^4 b_{123} + 12 E^2 b_{123}^2 - 33 E^3 c_{123} + 12 E b_{123} c_{123} + 8 c_{123}^2 \right] \\
\nonumber & \qquad + \frac{C_2}{2 E^2} \left[ -E^5 + 3 E^3 b_{123} + 4 E b_{123}^2 - 11 E^2 c_{123} + 4 b_{123} c_{123} \right] \\
& \qquad + 2  C_{TT} (p_1^3 + p_2^3 + p_3^3),
\label{tttformfac}
\end{align}
where $C_{TT}$ is the normalization of the 2-point function and the symmetric polynomials appearing are defined in \eqref{c123def}.
As discussed above, these form factors are defined up to the following degeneracies, derived in appendix A.5 of \cite{Bzowski:2017poo}:
\begin{align}
\nonumber \delta A_{1} &=f+f\left(p_{1}\leftrightarrow p_{3}\right)+f\left(p_{2}\leftrightarrow p_{3}\right)+g+g\left(p_{1}\leftrightarrow p_{3}\right)+g\left(p_{2}\leftrightarrow p_{3}\right)\\
\nonumber \delta A_{2} &=\left(p_{3}^{2}-p_{1}^{2}-p_{2}^{2}\right)f\\&\qquad \nn +p_{3}^{2}g+\frac{1}{2}\left(p_{1}^{2}-p_{2}^{2}+p_{3}^{2}\right)g\left(p_{1}\leftrightarrow p_{3}\right)+\frac{1}{2}\left(-p_{1}^{2}+p_{2}^{2}+p_{3}^{2}\right)g\left(p_{2}\leftrightarrow p_{3}\right)+h\\
\nonumber \delta A_{3}&=-\frac{1}{4}J^{2}f+p_{3}^{2}h\\
\nonumber \delta A_{4}&=\frac{1}{4}J^{2}g+\frac{1}{2}\left(p_{1}^{2}+p_{2}^{2}-p_{3}^{2}\right)\left(h\left(p_{1}\leftrightarrow p_{3}\right)+h\left(p_{2}\leftrightarrow p_{3}\right)\right)\\
 \delta A_{5}&=\frac{1}{4}J^{2}\left(h+h\left(p_{1}\leftrightarrow p_{3}\right)+h\left(p_{2}\leftrightarrow p_{3}\right)\right),
\label{tttdeg3}
\end{align}
Here, $f$, $g$ and $h$ are symmetric under $p_1\leftrightarrow p_2$, but are otherwise arbitrary functions of the momentum magnitudes.  
Unless otherwise specified, the ordering of arguments is assumed to be $f=f(p_1,p_2,p_3)$.  
Our aim is now to use these degeneracies to expose the underlying amplitude structure in the correlators.  
For these purposes, it will in fact  be sufficient to use an $h$ that is fully symmetric under permutations of the momenta.

First, as noted in \cite{Bzowski:2013sza, Bzowski:2017poo}, we can use these degeneracies to set $C_2=0$. 
The required $f$, $g$ and $h$ can be found by eliminating the terms proportional to $C_2$ in $A_3$, $A_4$ and $A_5$.  With this choice, the $C_2$ dependence in $A_1$ and $A_2$ then cancels out.  Next,  we may choose a new $f$, $g$ and $h$ to get rid of the terms proportional to $C_1$ in $A_3$, $A_4$ and $A_5$.\footnote{The required $f$, $g$ and $h$ are listed in the accompanying {\sc Mathematica} file.} The correlator can now be re-expressed in the compact form 
\begin{align}\label{3dTTTres}
\left\langle TTT\right\rangle &=-\frac{960 \,C_{1}c_{123}^{2}}{J^{2}E^{5}}\left(\mathcal{A}_{W^{3}}+\frac{1}{2}E\mathcal{A}_{F^{3}}\tilde{\mathcal{A}}_{YM}\right)\nn\\[1ex]&\quad +2C_{TT} \left[\left(\frac{c_{123}}{E^{2}}+\frac{b_{123}}{E}-E\right)\mathcal{A}_{EG}+\left(p_{1}^{3}+p_{2}^{3}+p_{3}^{3}\right)\tilde{\mathcal{A}}_{contact}\right],
\end{align}
where\footnote{
We can eliminate this contact term by using the perturbed metric  
$g_{\mu\nu} = [e^\gamma]_{\mu \nu}$
 as done in cosmology \cite{Maldacena:2002vr}.
The new 3-point function defined by taking functional derivatives with respect to $\gamma_{\mu\nu}$  then satisfies  $\<TTT\>_{new} = \<TTT\> -2 C_{TT}(p_1^3+p_2^3+p_3^3) \tilde{\mathcal{A}}_{contact}$.
} 
\[\label{Acontact}
\tilde{\mathcal{A}}_{contact}=\epsilon_{1}\cdot\epsilon_{2}\,\epsilon_{2}\cdot\epsilon_{3}\,\epsilon_{3}\cdot\epsilon_{1}.
\]
While this expression is equal to \eqref{tttformfac} via the degeneracies, writing the correlator in this way makes the connection to scattering amplitudes completely manifest.  We see the $\phi R^2$ amplitude has dropped out while the others are given by  flat space limits consistent with \eqref{TTTflat1} and \eqref{TTTflat3}:
\begin{align}
\lim_{E\rightarrow0}\frac{E^{6}}{c_{123}}\left.\left\langle TTT\right\rangle \right|_{C_{TT}=0}&\propto \mathcal{A}_{W^{3}},\qquad
\lim_{E\rightarrow0}\frac{E^{2}}{c_{123}}\left.\left\langle TTT\right\rangle \right|_{C_{1}=0}\propto\mathcal{A}_{EG}.
\end{align}

The term in \eqref{3dTTTres} proportional to $C_1$ is analyzed further 
in appendix \ref{wittendiagrams},  where we show it can be derived from a 
$W^3$ interaction in dS$_4$.  Relating this to a flat space Feynman diagram without energy conservation, we find that 
\[
\mathcal{A}_{W^{3}}+\frac{1}{2}E\mathcal{A}_{F^{3}}\tilde{\mathcal{A}}_{YM}
\propto 
\frac{J^{2}}{c_{123}E}\,\mathcal{M}_{W^{3}}
\]
so the term proportional to $C_1$ in \eqref{3dTTTres} can be written in the form $c_{123}\mathcal{M}_{W^{3}}/E^{6}$ where

\begin{equation}
\mathcal{M}_{W^{3}}\propto W_{\mu\nu\rho\lambda}^{(1)}W_{\rho\lambda\omega\sigma}^{(2)}W_{\omega\sigma\mu\nu}^{(3)},\qquad W_{\mu\nu\rho\lambda}^{(\alpha)}=p_{[\mu}^{(\alpha)}\epsilon_{\nu]}^{(\alpha)}p_{[\rho}^{(\alpha)}\epsilon_{\lambda]}^{(\alpha)}.
\label{w3}
\end{equation}
Here, repeated indices should again be contracted with the $(3+1)$-dimensional Minkowski metric, and the momenta and polarization vectors are those given in \eqref{bulkpdef} and \eqref{epdef} with $(\alpha)$ labelling the insertion.  Moreover, the two possible index contractions of $W^3$ are equivalent as we show in appendix \ref{wittendiagrams}, so we can further relate $\mathcal{M}_{W^3}$ to $\mathcal{M}_{F^3}$ defined in  \eqref{MF3res}:
\[
\mathcal{M}_{F^{3}} \propto M_{\mu\mu}, \qquad \mathcal{M}_{W^{3}} \propto  M_{\mu\nu}M_{\nu\mu}, \qquad M_{\mu\nu} =  F_{\mu\rho}^{(1)}F_{\rho\sigma}^{(2)}F_{\sigma\nu}^{(3)}.
\]
This relation suggests the existence of a double copy structure beyond the flat space limit.

\subsection{\texorpdfstring{$\langle JJ\O\rangle$}{<JJO>}}

In $d=3$, the relevant form factors  for two currents and a marginal scalar are \cite{Bzowski:2013sza,Bzowski:2018fql}
\begin{align}
A_1 = \frac{C_{1}}{E^{2}}(E+p_{3}), \qquad 
A_2 = -\frac{C_1}{2E}(E-2p_3)(E+p_3).
\end{align}
Plugging this into \eqref{jjogdcor} then gives 
\[\label{JJO3dres}
\left\langle JJ\O\right\rangle =-\frac{C_{1}}{2E^2}(E+p_3)\Big(2\,
\epsilon_{1}\cdot p_{2}\,\epsilon_{2}\cdot p_{1}+E(E-2p_3)\epsilon_{1}\cdot\epsilon_{2}\Big).
\]
As there are no degeneracies for this correlator in $d=3$, the flat space limit is the same as given earlier in \eqref{JJOflat1} and \eqref{JJOflat2}.  
In appendix \ref{wittendiagrams}, we show that 
\[\label{MF2def}
\<JJ\O\> \propto \frac{(E+p_3)}{E^2} \mathcal{M}_{\phi F^2}, \qquad
\mathcal{M}_{\phi F^2} = F^{(1)}_{\mu\nu}F^{(2)}_{\mu\nu},
\]
where $F^{(\alpha)}_{\mu\nu}$ as given in \eqref{MF3res}.

\subsection{\texorpdfstring{$\left\langle TT\O\right\rangle$}{<TTO>}}

As we discussed in section \ref{amps},
multiplying a YM amplitude with an $F^3$ amplitude gives rise to a nonzero graviton-graviton-scalar amplitude in the $\phi R^2$ theory (see \eqref{phiw2}). We will now demonstrate that this amplitude arises in the flat space limit of a $d=3$ correlator with two stress tensors and a marginal scalar. 

The relevant form factors are \cite{Bzowski:2013sza, Bzowski:2018fql}
\begin{align}
\nonumber A_1 &= \frac{C_1}{E^4} \mathcal{E}_1,
\\
\nonumber A_2 &=\frac{C_1}{E^3} \big( - \mathcal{E}_1 (E - 2p_3) + 2 \mathcal{E}_2 b_{12} \big),
\\
A_3 &=\frac{C_1}{4 E^2}(E-2p_3) \big( \mathcal{E}_1 (E -2 p_3) - 4 \mathcal{E}_2 b_{12} \big),
\end{align}
with 
\begin{align}
\mathcal{E}_1 & = E^3 + E b_{123} + 3 c_{123}, \qquad
\mathcal{E}_2  =E^2+p_3(E-p_3)
\end{align}
and symmetric polynomials as defined in \eqref{c123def}.
As with $\ang{TTT}$, these form factors are defined up to the degeneracy
\begin{equation}
\delta A_{1}=F,\qquad \delta A_{2}=-(p_{1}^{2}+p_{2}^{2}-p_{3}^{2})F,\qquad A_{3}=-\frac{1}{4}J^{2}F,
\label{ttodeg}
\end{equation}
where $F=F(p_1,p_2,p_3)$ is symmetric under $p_1\leftrightarrow p_2$ but otherwise an arbitrary function of momentum magnitudes.\footnote{See  (3.154) of \cite{Bzowski:2018fql}.}
Choosing
\[
F = -\frac{C_1}{E^3}\,\mathcal{E}_2,
\]
and inserting the results into \eqref{ttogdcor}, we obtain
\begin{equation}
\left\langle TT\O\right\rangle =C_1 b_{12} \frac{(E+3p_3)}{E^4} \,\big(2\,\ep_1\cdot p_2\,\ep_2\cdot p_1+E(E-2p_3)\, \ep_1\cdot\ep_2
\big)^2.
\label{phiw2tt}
\end{equation}
An equivalent spinor version is given in \eqref{spinorresults} of appendix \ref{spinors}.
 From either expression, it is clear that the flat space limit is given by
\[
\lim_{E\rightarrow0}\frac{E^{4}}{c_{123}}\left\langle TT\O\right\rangle \propto\mathcal{A}_{\phi R^{2}}^{220}
\]
where $\mathcal{A}_{\phi R^{2}}^{220}$ is the prediction of the double copy in \eqref{phiw2} and \eqref{phir2vec}.  

The double copy structure of \eqref{phiw2tt} is discussed further in appendix \ref{wittendiagrams}, where we show it can be derived from a $\phi W^2$ interaction in dS$_4$.  Relating this to a flat space Feynman integral without energy conservation, we find
\[
\left\langle TT\O\right\rangle \propto \frac{b_{12}(E+3p_3)}{E^4}\, (\mathcal{M}_{\phi F^2})^2\propto \frac{b_{12}(E+3p_{3})}{(E+p_{3})^{2}} \left\langle JJ\O\right\rangle^2,
\]
with $\mathcal{M}_{\phi F^2}$ as given in \eqref{MF2def}.
Again, this result suggests the existence of a double copy structure beyond the flat space limit.

\section{Conclusion} \label{conclusion}

It is becoming clear that there are deep connections between scattering amplitudes and correlation functions in quantum field theory. In this paper, we demonstrate that the double copy structure relating gauge and gravitational scattering amplitudes is encoded in the 3-point correlators of currents, stress tensors and marginal scalar operators in general CFTs. Starting with the general solutions to the conformal Ward identities in momentum space, we first dressed them with polarization vectors. For correlation functions in odd dimensions, we then derived a simple formula for the leading behavior of triple-$K$ integrals in the flat space limit, and showed that in this limit the correlators are spanned by certain scattering amplitudes in one higher dimension which are related via a double copy.

For $d=3$, the analysis is more subtle since the stress tensor correlators exhibit degeneracies. Using these degeneracies to expose the underlying amplitude structure of the correlation functions leads to considerable simplifications. For example, we show that the $\phi R^2$ contribution to stress tensor correlators vanishes. Instead, the double copy predicts that  the graviton-graviton-scalar  amplitude of this theory should arise from a correlator of two stress tensors and a marginal scalar. Moreover, we show that $d=3$ correlators can be written in terms of $(3+1)$-dimensional flat space amplitudes without energy conservation, suggesting double copy structure beyond the flat space limit.  We also obtain concise formulae for the correlators in terms of spinor-helicity variables.

There are a number of interesting questions to explore: 
\begin{itemize}
\item In odd dimensions, the correlators we consider are rational functions in momentum space, but in even dimensions their analytic structure is more complicated making the flat space limit subtle to evaluate.
 In $d=4$, they can be reduced to a single master integral corresponding to a 1-loop triangle diagram in flat space 
 \cite{Bzowski:2015pba}. The analytic properties of this integral have been discussed in, {\it e.g.,} \cite{tHooft:1978xw, Davydychev:1995mq, Chavez:2012kn}, which should be useful for evaluating the flat space limit.  A double copy structure for the 3-point contribution from the Euler trace anomaly has also been found in  \cite{Bzowski:2017poo}.
 It would be interesting to understand the analytic structure of 3-point correlators in general even dimensions and to develop a method for extracting their flat space limit.  In $d=2$ there are further subtleties because conformal symmetry becomes enhanced to Virasoro symmetry. Futhermore, it may be possible to derive current correlators from Witten diagrams for two kinds of gauge theories in the bulk, notably 3d Yang-Mills and Chern-Simons-matter theories, both of which are known to square into gravitational theories in the flat space limit \cite{Bargheer:2012gv,Huang:2012wr}.

\item The simplicity of our results for $d=3$ correlators suggests an underlying worldsheet description for Witten diagrams in dS$_4$, similar to those developed for scattering amplitudes in flat space. Worldsheet formulae have recently been derived for gauge and gravitational amplitudes in plane-wave backgrounds \cite{Adamo:2017sze}, where a double copy was also implemented at the level of the classical background.\footnote{Double copies for classical backgrounds were first considered in \cite{Monteiro:2014cda} and were explored for maximally symmetric backgrounds in \cite{Carrillo-Gonzalez:2017iyj}.} In the context of CFT correlators, however, it would be desirable to have worldsheet descriptions for both gauge and gravitational Witten diagrams in de Sitter space. If a worldsheet description of 3-point Witten diagrams can be defined, the main challenge would then be to extend this description to higher points by incorporating a curved space analogue of the scattering equations \cite{Adamo:2014wea}, which could provide a useful tool for studying holography in the supergravity approximation.

\item It has recently been shown that for a class of conformally coupled scalar theories in dS$_4$, the wavefunction of the universe can be expressed in terms of volumes of polytopes \cite{Arkani-Hamed:2017fdk}. Ultimately, these can be derived from flat space Feynman diagrams with external propagators ending on a fixed timeslice. Since we have found a similar structure for CFT correlators in three dimensions, it would be interesting to explore if they also have a polytope interpretation \cite{Arkani-Hamed:2018ign}.
In the flat space limit, the worldsheet and polytope descriptions of scattering amplitudes are intimately connected \cite{Spradlin:2009qr,Dolan:2009wf,Nandan:2009cc,ArkaniHamed:2009dg,Arkani-Hamed:2013jha,Farrow:2017eol,Arkani-Hamed:2017mur}, so it would be interesting to see if this connection extends to dS$_4$ background.

\item Since 3-point amplitudes are the building blocks for all higher-point scattering amplitudes via BCFW recursion and unitarity, and 3-point correlators are the building blocks for higher-point CFT correlators via the OPE, we expect that our results can be extended to higher points (see {\it e.g.,} \cite{Raju:2012zr, Raju:2011mp, Raju:2010by,Raju:2012zs}). It would therefore be interesting to find general solutions to the conformal Ward identities for 4-point correlators in momentum space and see to what extent they have a double copy structure in the flat space limit. In the case of 3-point correlators, this can be anticipated from scattering amplitudes in bosonic string theory, so it would be interesting to see if this continues to hold at higher points. 
\end{itemize}

We hope to report on these directions in the future.

\acknowledgments{
We thank Paul Heslop for discussion. JF is funded by EPSRC PhD scholarship EP/L504762/1, AL is supported by the Royal Society as a Royal Society University Research Fellowship holder, and PM is supported by the STFC through an Ernest Rutherford Fellowship. 
}

\appendix

\section{de Sitter correlators} \label{wittendiagrams}

In this appendix, we explore further the double copy structure of CFT correlators by relating them to 3-point functions in de Sitter spacetime.
For three-dimensional CFTs, certain of these correlators have the special property that they further reduce to Feynman diagrams in $(3+1)$-dimensional {\it flat space}, without energy conservation, dressed by overall conformal time integrals.  This clarifies the origin of the double copy structure, and moreover implies its persistence beyond leading order in the flat space limit.

\begin{figure}
\centering
	       \includegraphics[scale=1]{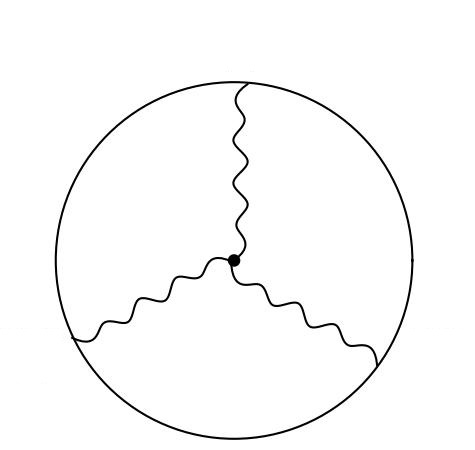}\hspace{1cm} \includegraphics[scale=1]{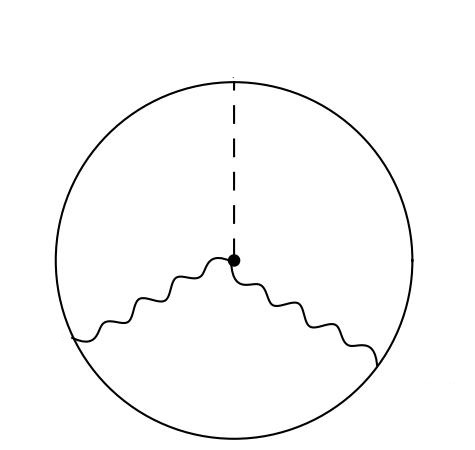}
    \caption{Tree-level diagrams for 3-point correlators. The first diagram describes $W^3$ and $F^3$ interactions, while the second diagram describes $\phi F^2$ and $\phi W^2$ interactions.} 
    \label{3pt}
\end{figure}

We choose coordinates for the de Sitter background metric
\[
\mathrm{d}s^{2}=\eta^{-2}\left(-\mathrm{d}\eta^{2}+\mathrm{d} x_i^2\right),
\]
where $-\infty<\eta<0$ is the conformal time and $i=1,2,3$ labels the spatial coordinates.
Our focus will then be to evaluate the $F^3$, $W^3$, $\phi F^2$ and $\phi W^2$ diagrams contributing to $d=3$ $\<JJJ\>$, $\<TTT\>$, $\<JJ\O\>$ and $\<TT\O\>$ correlators respectively, as illustrated in figure \ref{3pt}.  
For a gauge field on dS$_4$, the on-shell wavefunction is simply a plane wave $A_\mu \propto e^{ip\eta}\ep_\mu$ as in flat space, so the reduction to Feynman diagrams comes as no surprise.  For a graviton on dS$_4$, however, the on-shell wavefunction is not a plane wave but instead has time-dependence $\gamma_{\mu\nu}\propto (1-ip\eta)e^{ip\eta}\ep_\mu\ep_\nu$. 
Nevertheless, as discussed in \cite{Maldacena:2011nz}, the 
linearized Weyl tensor is conformal to that for a graviton in flat space:
\begin{equation}
W^{(dS)}{}^{\mu}{}_{\nu\rho\sigma}\left[\left(1-ip\eta\right)e^{ip\eta}\right]=-ip\eta \,W^{(flat)}{}^\mu{}_{\nu\rho\sigma}\left[e^{ip\eta}\right].
\label{weyltransform}
\end{equation}
This allows us to perform all index contractions as if in $(3+1)$-dimensional Minkowski space using the momenta and polarization vectors given in \eqref{bulkpdef} and \eqref{epdef}.  The de Sitter background can then be restored by multiplying with an overall conformal time integral consisting of plane wave external states multiplied by additional factors of $\eta$ as appropriate.  These factors  derive from the de Sitter measure, the inverse metrics appearing in index contractions, and any conformal factors arising from \eqref{weyltransform}.  Further details of this approach may be found in \cite{Maldacena:2011nz},\footnote{Note our wavefunctions are the Bunch-Davies ones multiplied by a factor of $p^{3/2}$.   In holographic cosmology where the standard Bunch-Davies vacuum is assumed, these factors arise instead through products of the 2-point function in the holographic formulae for 3-point functions, see \cite{Maldacena:2002vr,McFadden:2010vh,McFadden:2011kk}.}
 where an alternative calculation of the $W^3$ diagram is provided.   (The result is equivalent to that here after using the degeneracies in  \eqref{tttdeg3}.)

We begin with the 3-point amplitude for the $F^3$ theory. 
Working in $(3+1)$-dimensional Minkowski spacetime, with momenta and polarizations as given in \eqref{bulkpdef} and \eqref{epdef}, we have
\[
\mathcal{M}_{F^{3}}\propto F_{\mu\nu}^{(1)}F_{\nu\rho}^{(2)}F_{\rho\mu}^{(3)},\qquad F_{\mu\nu}^{(\alpha)}=p_{[\mu}^{(\alpha)}\epsilon_{\nu]}^{(\alpha)},
\]
where the superscript $(\alpha)$ labels the insertions, and in particular  
\[
F_{0i}^{(\alpha)}=\frac{1}{2}p^{(\alpha)}\epsilon_{i}^{(\alpha)},\qquad F_{ij}^{(\alpha)}=p_{[i}^{(\alpha)}\epsilon_{j]}^{(\alpha)}.
\]
Evaluating the contractions, we find
\begin{align}
\mathcal{M}_{F^{3}} \propto 2\mathcal{A}_{F^3}+E\tilde{\mathcal{A}}_{YM},
\end{align}
with $\tilde{\mathcal{A}}_{YM}$ as given in \eqref{tildeAYMdef}.
To return to dS$_4$, we  dress this flat-space amplitude with the conformal time integral
\[
{\rm Im}\left(\int_{-\infty}^{0}\mathrm{d}\eta\,\eta^{2}e^{iE\eta}\right)\propto \frac{1}{E^{3}}.
\]
As expected, we thus recover the term proportional to  $C_1$ in \eqref{F3full}.  
Here, the factor of $\eta^2$ derives from the measure $\sqrt{-g}$ and three inverse metrics, and for convergence we infinitesimally rotate the contour by sending $\eta\rightarrow (1-i\ep)\eta$.
The need for taking the imaginary part follows from the in-in formalism, see {\it e.g.,} \cite{Maldacena:2002vr}.

Next let us consider the $W^3$ amplitude, which in flat space is given by 
\begin{equation}
\mathcal{M}_{W^{3}}\propto W_{\mu\nu\rho\sigma}^{(1)}W_{\rho\sigma\lambda\omega}^{(2)}W_{\lambda\omega\mu\nu}^{(3)},\qquad W_{\mu\nu\rho\sigma}^{(\alpha)}=p_{[\mu}^{(\alpha)}\epsilon_{\nu]}^{(\alpha)}p_{[\rho}^{(\alpha)}\epsilon_{\sigma]}^{(\alpha)}.
\label{W3}
\end{equation}
Here, the linearized Weyl tensor reduces to the linearized Riemann tensor, which for the transverse traceless graviton can be written as the tensor product of two Yang-Mills field strengths.  The contractions are thus equivalent to evaluating
\[
(F^{(1)}_{\rho\sigma}F^{(2)}_{\rho\sigma})(F^{(2)}_{\lambda\omega}F^{(3)}_{\lambda\omega})(F^{(3)}_{\mu\nu}F^{(1)}_{\mu\nu}).
\]
Since 
\begin{align}\label{F2contr}
F^{(1)}_{\rho\sigma}F^{(2)}_{\rho\sigma} \propto
2\,\ep_1\cdot p_2\,\ep_2\cdot p_1+E(E-2p_3)\, \ep_1\cdot\ep_2,
\end{align}
we  can now write $\mathcal{M}_{W^3}$ in the form \eqref{tttgeneral}.  Up to an overall constant of proportionality, the corresponding form factors are  
\begin{align}
&A_1 = 8, \quad A_2 = 4E(2p_3-E), \quad A_3 = 0,\quad A_4 = 2E^2(E-2p_1)(E-2p_1),\quad  A_5 = -J^2 E^2.
\end{align}
Applying the degeneracy \eqref{tttdeg3} with
\[
f = \frac{16 E^2 p_3^2}{3J^2},\qquad g=\frac{8E^2}{3 J^2}\big(p_1^2+p_2^2-6p_1p_2-p_3^2\big), \qquad h=\frac{4E^2}{3},
\]
we can further set $A_4$ and $A_5$ to zero whereupon 
\[\label{MW3eval}
\mathcal{M}_{W^{3}}\propto \frac{c_{123}E}{J^2} \left(\mathcal{A}_{W^{3}}+\frac{1}{2}E\mathcal{A}_{F^{3}}\tilde{\mathcal{A}}_{YM}\right).
\]
To return to dS$_4$, these form factors are dressed by the conformal time integral 
\[
c_{123}\,{\rm Im}\left(\int_{-\infty}^0 \mathrm{d}\eta\,\eta^{5}e^{iE\eta}\right)\propto \frac{ c_{123}}{E^{6}},
\]
yielding the expected term proportional to $C_1$ in \eqref{3dTTTres}.
Here, the factor of $c_{123} \,\eta^5$ in the conformal time integral derives from the measure, three copies of the conformal factor in \eqref{weyltransform} and three inverse metrics. 

Since $\mathcal{A}_{W^3} = (\mathcal{A}_{F^3})^2$, we can pull out a factor of $\mathcal{A}_{F^3}$  in \eqref{MW3eval} obtaining a product structure even for $E\neq 0$. Moreover, noting that in $(3+1)$-dimensions 
\[
0 = W^{(1)}{}^{\mu\nu}{}_{[\mu\nu}W^{(2)}{}^{\rho\sigma}{}_{\rho\sigma}W^{(3)}{}^{\lambda\omega}{}_{\lambda\omega]},
\]
we can equivalently 
re-write the index contractions in \eqref{W3} as
\[
\mathcal{M}_{W^{3}}\propto W_{\mu\nu\rho\lambda}^{(1)}W_{\rho\sigma\mu\omega}^{(2)}W_{\lambda \omega\nu\sigma}^{(3)}.
\]
We then obtain the manifest square
\[
\mathcal{M}_{F^{3}} \propto M_{\mu\mu}, \qquad \mathcal{M}_{W^{3}} \propto  M_{\mu\nu}M_{\nu\mu}, \qquad M_{\mu\nu} =  F_{\mu\rho}^{(1)}F_{\rho\sigma}^{(2)}F_{\sigma\nu}^{(3)}.
\]
Evaluating the index contractions from this starting point is more involved, but the result is equivalent to \eqref{MW3eval} after using the degeneracies.

From the above calculations, we can easily compute the diagram corresponding to a $\phi F^2$ interaction. In particular, in flat space this is just given by
\[
\mathcal{M}_{\phi F^{2}} \propto F_{\mu\nu}^{(1)}F_{\mu\nu}^{(2)},
\]
which was computed in \eqref{F2contr}. Dressing this with the conformal time integral
\[
{\rm Im}\left(\int_{-\infty}^0 \mathrm{d}\eta\,(1-ip_{3}\eta)e^{iE\eta}\right)\propto \frac{E+p_{3}}{E^{2}},
\]
then gives \eqref{JJO3dres}.  Here, the factors of $\eta$ from the measure and two inverse metrics cancel out, while the remainder is the on-shell wavefunction for a massless scalar in dS$_4$.

Finally, let us consider the $\phi W^2$ interaction.  Starting again in flat space, we have 
\begin{equation}
\mathcal{M}_{\phi R^{2}}^{220}\propto W_{\mu\nu\rho\sigma}^{(1)}W_{\mu\nu\rho\sigma}^{(2)}
\label{w2p}
\end{equation}
The index contractions are then equivalent to
\[
(F^{(1)}_{\mu\nu}F^{(2)}_{\mu\nu})^2,
\]
which is the square of \eqref{F2contr}.
To return to dS$_4$, we dress this amplitude with
 the conformal time integral
\[
p_{1}p_{2}\,{\rm Im}\left(\int_{-\infty}^0 \mathrm{d}\eta\,\eta^{2}(1-ip_{3}\eta)e^{iE\eta}\right)\propto b_{12}\frac{(E+3p_3)}{E^4}
\]
recovering \eqref{phiw2tt}.
Here, we obtained a factor of $p_1 p_2 \eta^2$ from the measure along with two copies of \eqref{weyltransform} and two inverse metrics,
and the remaining factor is  the on-shell wavefunction for a massless scalar in dS$_4$.

\section{Spinor helicity formalism} \label{spinors}

In this appendix we describe the spinor helicity formalism of \cite{Maldacena:2011nz} for $d=3$ CFT correlators, or equivalently cosmological correlators in dS$_4$. As an application of the formalism, we verify the degeneracies of stress tensor correlators, and we write the 3d correlators considered in this paper in terms of spinor variables. We also present various useful identities for manipulating these variables, which are also implemented in the {\sc{Mathematica}} file {\tt{dSDoubleCopy.nb}}.

The $d=3$ spinor helicity formalism can be derived from $(3+1)$ dimensions by introducing a vector $\tau^\mu = (1,0,0,0)$ \cite{Lipstein:2012kd}. In terms of spinor indices this is given by $\tau^{\dot{\alpha}\beta}=\epsilon^{\alpha\dot{\beta}}$. This vector can then be used convert dotted indices into undotted indices. Applying this to a 4d null momentum then gives
\[
\tau_{\dot{\alpha}}^{\beta}\lambda^{\alpha}\tilde{\lambda}^{\dot{\alpha}}=\lambda^{\alpha}\bar{\lambda}^{\beta}=p^{\alpha\beta}+\epsilon^{\alpha\beta}p,
\]
where $\bar{\lambda}^{\alpha} = \tau^{\alpha}_{\dot{\alpha}}\tilde{\lambda}^{\dot{\alpha}}$, $p^{\alpha\beta}=\lambda^{(\alpha}\tilde{\lambda}^{\beta)}$ is the 3d momentum and $p=\frac{1}{2}\langle \lambda\bar{\lambda}\rangle$ is its magnitude (where spinor brackets were defined in section \ref{amps}). The spinors can be explicitly parametrized in terms of 3-momenta according to
\[
\lambda^\alpha(\bs{p}) =  \left(
\begin{array}{c} {\sqrt{|\bs{p}| + \bs{p}^3}} \\[1ex]\dfrac{\vec{p}^1 - i \bs{p}^2}{\sqrt{|\bs{p}| + \bs{p}^3}}\end{array}\right)
\]
Note that our conventions differ from those of \cite{Maldacena:2011nz} by a factor $\sqrt{2}$ in the spinors. The $\tilde{\lambda}^{\dot{\alpha}}$ spinor is then the complex conjugate of this expression, and the $\bar{\lambda}^{\alpha}$ spinor is calculated by multiplying by $\tau^\alpha_{\dot{\alpha}}$.

A spatial 3-momentum can be lifted to a null 4-momentum given by $\lambda^{\alpha}\tilde{\lambda}^{\dot{\alpha}}$, and given a set of $n$ 3-momenta which satisfy momentum conservation, their associated 4-momenta will not in general satisfy energy conservation:
\[
\sum_{i=1}^n \tilde{\lambda}_i^{\dot{\alpha}} \lambda_i^\beta = \tau^{\dot{\alpha}\beta} \sum_{i=1}^n p_i = \tau^{\dot{\alpha}\beta} E
\]
These 4-momenta are naturally interpreted as null momenta in dS$_4$, since energy is not conserved in this background. This leads to various useful identities in terms of the spinor variables at three points. We find that at three points, any product of two spinor brackets which is little group invariant in one or two particle labels can be reduced to a function of the momentum magnitudes only. The full set of identities producing this kind of reduction is given by
\begin{align}
\label{eq:3ptmomcon}
\ang{i\bar{i}} &= 2 p_i\nn \\
\ang{j\bar{i}}\ang{ik} &= \ang{jk}(p_k-p_i-p_j) \nn\\
\ang{j\bar{i}}\ang{i\bar{k}} &= \ang{j\bar{k}}(p_k-p_i+p_j) \nn\\
\ang{ji}\ang{\bar{i}\bar{k}} &= \ang{\bar{i}\bar{k}}(p_k+p_i+p_j) = \ang{\bar{i}\bar{k}} E \nn\\
\ang{j\bar{i}}\ang{i\bar{j}} &= (p_k-p_j+p_i)(p_k-p_i+p_j) \nn\\
\ang{ji}\ang{\bar{i}\bar{j}} &= p_k^2 - (p_i+p_j)^2.
\end{align}

Polarization vectors can also be expressed in terms of spinor variables. Fixing a gauge such that the time component of the vectors is zero, we find the following formulae
\begin{align}
\epsilon_-^{\alpha\beta} = \frac{\lambda^\alpha\lambda^\beta}{p_i}, \qquad
\epsilon_+^{\alpha\beta} = \frac{\bar{\lambda}^\alpha\bar{\lambda}^\beta}{p_i}.
\end{align}
Dot products between momenta and spinors are then given by
\begin{align}
\label{eq:poltospin}
2 p_i\cdot p_j &= \ang{ij}\ang{\bar{j}\bar{i}}, \qquad
2 p_i\cdot \epsilon_j^- = \frac{\ang{\bar{i}j}\ang{ji}}{p_j},\qquad
2 p_i\cdot \epsilon_j^+ = \frac{\ang{\bar{i}\bar{j}}\ang{\bar{j}i}}{p_j},\nonumber\\
 2\epsilon_i^-\cdot \epsilon_j^- &= -\frac{\ang{ij}^2}{p_i p_j},\qquad
 2\epsilon_i^+\cdot \epsilon_j^- = -\frac{\ang{\bar{i}j}^2}{p_i p_j},\qquad
 2\epsilon_i^+\cdot \epsilon_j^+ = -\frac{\ang{\bar{i}\bar{j}}^2}{p_i p_j}.
\end{align}

As a first application, let us verify that the degeneracies used to simplify stress tensor correlators in section \ref{3d}. Since the $\left\langle TTT\right\rangle$ degeneracies are constructed as sums over permutations of the $\left\langle TT\O\right\rangle$ degeneracy, it is sufficient to consider the $\left\langle TT\O\right\rangle$ degeneracy without loss of generality. The $\left\langle TT\O\right\rangle$ degeneracy in \eqref{ttodeg} can be derived by adding the following object (multiplied by an arbitrary function of the momentum magnitudes) to the correlator
\[
D = (\epsilon _1\cdot p_2)^2 \,(\epsilon _2\cdot p_3)^2-\left(p_1^2+p_2^2-p_3^2\right) \epsilon _1\cdot \epsilon _2 \,\epsilon _1\cdot p_2\, \epsilon _2\cdot p_3 -\frac{1}{4} J^2 (\epsilon _1\cdot \epsilon _2)^2, 
\]
and decomposing it in terms of form factors according to \eqref{ttogdcor}. Our task is then to show that this object vanishes, and we can do so by writing it in terms of spinors. Indeed, for $--$ and $-+$ helicity assignments we find that
\begin{align}
D^{--} &= \frac{\langle 12\rangle ^4 \left(2 \left(p_1^2+p_2^2-p_3^2\right) \langle 1\bar{2}\rangle  \langle 2\bar{1}\rangle +\langle 1\bar{2}\rangle ^2 \langle 2\bar{1}\rangle ^2-J^2\right)}{16 p_1^2 p_2^2} , \\[2ex]
D^{-+} &=-\frac{\langle 1\bar{2}\rangle ^4 \left(2 \left(p_1^2+p_2^2-p_3^2\right) \langle 12\rangle  \langle \bar{1}\bar{2}\rangle -\langle 12\rangle ^2 \langle \bar{1}\bar{2}\rangle ^2+J^2\right)}{16 p_1^2 p_2^2},
\end{align}
both of which are identically zero on support of the identities given above.

It is now algorithmic using the relations between dot products and spinor brackets~(\ref{eq:poltospin}), and the identities~(\ref{eq:3ptmomcon}) to take any 3-point correlator in terms of polarization vectors and momenta and produce a function of spinor brackets. An explicit realization of these algorithms is given in the accompanying {\sc{Mathematica}} file, and calculations are shown for $\ang{TTT}$, $\ang{JJJ}$, $\<JJ\O\>$ and $\ang{TT\O}$. 

The results for a full set of different helicity assignments are
{\allowdisplaybreaks{
\begin{align}
\ang{T^-T^-T^-} &= \langle 12\rangle ^2 \langle 23\rangle ^2 \langle 31\rangle ^2\left(
C_1 \frac{120 c_{123}}{E^6}-C_3\frac{E^3 -E b_{123}-c_{123}+2 \left(p_1^3+p_2^3+p_3^3\right)}{8 c_{123}^2}\right), \nn\\[.3cm]
\ang{T^-T^-T^+} &= 
\langle 12\rangle ^2 \langle 2\bar{3}\rangle ^2 \langle \bar{3}1\rangle ^2 C_3 \frac{ \left(p_1+p_2-p_3\right){}^2\left(E^3-E b_{123}-c_{123}\right) -2E^2\left(p_1^3+p_2^3+p_3^3\right)}{8 E^2 c_{123}^2}
\nn\\[0.3cm]
&\hspace{-1.7cm}=\frac{\langle 12\rangle ^8}{\ang{12}^2\langle 23\rangle ^2 \langle 31\rangle ^2} \frac{C_3  J^4 }{8 E^4 c_{123}^2 } \Big(E^3-E b_{123}-c_{123}
- 2E^2\frac{\left(p_1^3+p_2^3+p_3^3\right)}{\left(p_1-p_2+p_3\right){}^2\left(-p_1+p_2+p_3\right){}^2}\Big),\nn \\[.3cm]
\ang{J^-J^-J^-} &=
\langle 12\rangle  \langle 23\rangle  \langle 31\rangle  \left(\frac{C_2 }{2 c_{123}}-  \frac{2C_1}{E^3} \right), \nn\\[.3cm]
\ang{J^-J^-J^+}
 &=\langle 12\rangle  \langle 2\bar{3}\rangle  \langle \bar{3}1\rangle C_2 \frac{ (p_1+p_2-p_3) }{2 E c_{123}} 
 =\frac{\langle 12\rangle ^4}{\ang{12}\langle 23\rangle \langle 31\rangle} \frac{C_2 J^2}{2 E^2 c_{123}}, \nn \\[.3cm]
\ang{J^-J^-\O} &=\langle12\rangle^{2}C_{1}\frac{\left(E+p_{3}\right)}{E^{2}},\nn \\[.3cm]
\ang{J^-J^+\O} &= 0 \nn \\[.3cm]
\ang{T^-T^-\O} &= \langle 12\rangle ^4 C_1 \frac{p_1 p_2 \left(E+ 3 p_3\right)}{E{}^4},\nn \\[.3cm]
\ang{T^-T^+\O} &= 0. \label{spinorresults}
\end{align}}}
Note that combining the identities~(\ref{eq:3ptmomcon}) together we recover the following relation, which is used to give different forms for $--+$ correlators,
\[
\frac{ \langle 12\rangle ^4}{\ang{12}\langle 23\rangle  \langle 31\rangle } \left(p_1-p_2-p_3\right) \left(p_1-p_2+p_3\right) = \ang{12}\langle 2\bar{3}\rangle \langle \bar{3}1\rangle . 
\]
Up to this relation, the spinor bracket structure of all 3-point correlators is fixed by little group covariance, and only the function of the momentum magnitudes $p_i$ needs to be calculated from the solution to the conformal Ward identities.

\bibliographystyle{JHEP}
\bibliography{dsamp_published}

\end{document}